\documentclass[final,3p,twocolumn,sort&compress]{elsarticle}

\usepackage{amsmath}
\usepackage{amssymb}
\usepackage[figurename=Fig.,labelsep=period]{caption}
\usepackage[colorlinks,citecolor=blue,urlcolor=blue,linkcolor=blue]{hyperref}

\begin{document}

\begin{frontmatter}



\title{A differential relation between the energy and electric charge of a dyon}

\author[tusur,tpu]{A.Yu.~Loginov}
\ead{a.yu.loginov@tusur.ru}

\address[tusur]{Tomsk State University of Control Systems and Radioelectronics, 634050 Tomsk, Russia}
\address[tpu]{Tomsk Polytechnic University, 634050 Tomsk, Russia}

\begin{abstract}
The differential relation between  the  energy and electric charge of a dyon is
derived.
The  relation expresses  the  derivative  of  the  energy  with  respect to the
electric charge in terms of the  boundary  value  for the temporal component of
the dyon's electromagnetic potential.
The use of  the  Hamiltonian  formalism  and  transition  to  the unitary gauge
make it possible  to  show  that  this  derivative is proportional to the phase
frequency of the electrically charged  massive  gauge fields forming the dyon's
core.
It follows from the differential relation  that  the energy and electric charge
of the non-BPS dyon cannot be arbitrarily large.
Finally, the dyon's properties are investigated numerically at different values
of the model parameters.
\end{abstract}


\end{frontmatter}

\section{Introduction}
\label{seq:I}

Electrically   charged   solitons    exist    in    both  $(2 + 1)$-dimensional
\cite{paul, khare_rao_227,  hong,  jw1,  jw2,  bazeia_1991, ptz_plb_339, ghosh,
arth_tchr_prd_54, deshaies_2006, lgn_2014,navarro_2017} and $(3+1)$-dimensional
\cite{jz_1975, prs_1975, cpns_1975, bkt_1999, rduchr_2006, klee, lee_yoon_1991,
ardoz_2009, tamaki_2014,brihaye_prd_2014,gulamov_2014,gulamov_2015,lshnir_2019,
loginov_2020} gauge field models.
The $(2 + 1)$-dimensional field models permitting the existence of electrically
charged solitons include the Chern--Simons gauge term \cite{JT,schonfeld, DJT},
and  therefore  their  gauge  fields  are  topologically  massive, resulting in
the short-range electric field.
In contrast, the  three-dimensional  electrically  charged  solitons  possess a
long-range  electric field because  the corresponding $(3+1)$-dimensional field
models include  only  the  Maxwell  gauge  term,  leading to the massless gauge
fields.

The three-dimensional  electrically  charged  solitons  can be both topological
\cite{jz_1975, prs_1975, cpns_1975, bkt_1999, rduchr_2006}  and  nontopological
\cite{klee, lee_yoon_1991,ardoz_2009,tamaki_2014,brihaye_prd_2014,gulamov_2014,
gulamov_2015, lshnir_2019, loginov_2020} type.
The properties of these two  types  of  solitons  are  substantially different.
The  existence of the electrically charged nontopological  solitons  is  due to
the presence of the conserved  Noether (electric) charge  and  the special form
of the self-interaction potential of scalar fields.
The  basic  property  of  the   nontopological   soliton   is  that  its  field
configuration is a stationary  (saddle  or  minimum)  point of the total energy
functional at a given fixed  value  of  the Noether charge \cite{fried, fried1,
fried2}.
This property results in the differential relation between  the  energy and the
Noether charge of  the  nontopological  soliton,  which,  in turn, determines a
number of the soliton's properties. 

On the  other  hand,  the  existence  of  the  topological  solitons, including
the electrically charged ones, is due to the topological nontriviality of their
field  configurations, which prevents  the  transitions of topological solitons
into the states with the lower energy.
In particular, the presence  of  a  potential term is not a necessary condition
for the existence of topological solitons \cite{Manton}.
The best known example of the three-dimensional topological solitons possessing
an electric charge is the dyon  solution \cite{jz_1975} of  the Georgi--Glashow
model \cite{GG_1972}.
In this Letter we  derive  the  differential  relation  between  the energy and
electric charge of this dyon solution.
We also ascertain that the differential  relation  determines  a  number of
properties of the dyon.
In particular, we show  that  it  does  not  allow the existence of the non-BPS
dyons possessing the arbitrary large electric charge and energy.


\section{Lagrangian and field equations of the model} \label{seq:II}

The Lagrangian density of the Georgi--Glashow model is
\begin{equation}
\mathcal{L}=-\frac{1}{4}F_{\mu \nu }^{a}F^{a\,\mu \nu }+\frac{1}{2}\left(
D_{\mu }\phi ^{a}\right) \left( D^{\mu }\phi ^{a}\right) - V\left( \phi
\right),                                                           \label{II:1}
\end{equation}
where
\begin{equation}
F_{\mu \nu }^{a}=\partial _{\mu }A_{\nu }^{a} - \partial _{\nu }A_{\mu
}^{a}-e\epsilon^{abc}A_{\mu }^{b}A_{\nu }^{c}                     \label{II:2}
\end{equation}
is the non-Abelian field strength,
\begin{equation}
D_{\mu}\phi^{a} = \partial_{\mu}\phi^{a} - e\epsilon^{abc}A_{\mu}^{b}\phi^{c}
                                                                   \label{II:3}
\end{equation}
is the  covariant  derivative  of  the  Higgs   field  $\boldsymbol{\phi}$, and
\begin{equation}
V\left(\phi\right) = \frac{\lambda}{4}\left(\phi^{a}\phi^{a} -
v^{2}\right)^{2}                                                   \label{II:4}
\end{equation}
is the self-interaction  potential  of  the  Higgs  field  $\boldsymbol{\phi}$.
In Eqs.~(\ref{II:2}) -- (\ref{II:4}),  $e$  is  the  gauge  coupling  constant,
$\lambda$ is the self-interaction coupling  constant  of  the Higgs  field, and
$v$ is the Higgs field vacuum expectation value.
The field equations of model (\ref{II:1}) have the form
\begin{eqnarray}
D_{\nu }F^{a\,\mu \nu }+e\epsilon^{abc}\phi ^{b}D^{\mu }\phi ^{c} & = & 0,
                                                                  \label{II:5a}
\\
D_{\mu }D^{\mu }\phi ^{a}+\lambda \left( \phi ^{b}\phi ^{b}-v^{2}\right)
\phi ^{a} &=& 0,                                                  \label{II:5b}
\end{eqnarray}
and the symmetric energy-momentum tensor is
\begin{flalign}
& T_{\mu \nu } = -F_{\mu \rho }^{a}F_{\nu }^{a\,\rho }+\left( D_{\mu }\phi
^{a}\right) \left( D_{\nu }\phi ^{a}\right)                        \label{II:6}
\\
& +\eta_{\mu \nu}\left[\frac{1}{4}F_{\rho \tau}^{a}F^{a\,\rho \tau}-
\frac{1}{2}\left(D_{\rho}\phi^{a}\right)\left(D^{\rho}\phi^{a}\right)
+V\left( \phi \right) \right]\!,                            \nonumber
\end{flalign}
where $\eta_{\mu \nu} = \text{diag}(+1,-1,-1,-1)$ is the metric tensor.

In model (\ref{II:1}), finite  energy  field  configurations  must  satisfy the
asymptotic   condition   $\underset{  r  \rightarrow  \infty} {\lim} \left\vert
\boldsymbol{\phi} \right\vert = v$, which is equivalent to  the  mapping of the
infinitely  distant  space   sphere   $S^{2}_{\infty}$  to  the  vacuum  sphere
$S_{\text{vac}}^{2}\!: \left\vert \boldsymbol{\phi} \right\vert = v$.
It is well  known  that  the  mappings $S^{2} \rightarrow S^{2}$ are split into
different topological classes, which are  characterised  by the integer winding
number $n$ according to the sphere's second homotopy group $\pi_{2}\left( S^{2}
\right) = \mathbb{Z}$.
In the topological sector with  the winding  number $n = 1$, model (\ref{II:1})
has the two well  known  topological  soliton  solutions: the 't~Hooft–Polyakov
monopole \cite{hooft_74, polyakov_74} and Julia--Zee dyon \cite{jz_1975}.
Both the 't~Hooft--Polyakov monopole and  Julia--Zee  dyon  possess the minimum
possible magnetic charge $g = 4\pi/e$.
At the same time, the  electric  charge  of the 't~Hooft--Polyakov  monopole is
equal to zero, whereas that of the dyon is nonzero.
When the electric charge tends to  zero,  the dyon field configuration smoothly
goes into the 't~Hooft--Polyakov monopole.

The field configuration of the dyon is  described  by the spherically symmetric
ansatz 
\begin{subequations}                                               \label{II:7}
\begin{eqnarray}
A^{a0}&=&n^{a}vj\left(r\right),                                   \label{II:7a}
\\
A^{ai} &=&\epsilon^{a i m}n^{m}\frac{1 - u\left(r\right)}{er},    \label{II:7b}
\\
\phi ^{a} &=&n^{a}vh\left( r\right),                              \label{II:7c}
\end{eqnarray}
\end{subequations}
where $n^{a}=x^{a}/r$.
We now  introduce  the  dimensionless  radial  variable $\rho = m_{V} r$, where
$m_{V} = e v$ is the mass of the electrically charged gauge bosons.
Then, the ansatz  functions $u\left(r\right)$, $j\left(r\right)$, and $h\left(r
\right)$  will  satisfy  the  system of nonlinear differential equations of the
second order:
\begin{flalign}
& u^{\prime \prime }\left( \rho \right) -\frac{u(\rho )\left( u(\rho
)^{2}-1\right) }{\rho ^{2}}                                       \label{II:8a}
\\
& -\left( h(\rho )^{2}-j(\rho )^{2}\right) u(\rho )=0, \nonumber
\\
& j^{\prime \prime }\left( \rho \right) +\frac{2}{\rho}j^{\prime}\left(\rho
\right) -\frac{2}{\rho ^{2}}u(\rho )^{2}j(\rho )=0,               \label{II:8b}
\\
& h^{\prime \prime }\left( \rho \right) +\frac{2}{\rho}h^{\prime}\left(\rho
\right) -\frac{2}{\rho ^{2}}u(\rho )^{2}h(\rho )                  \label{II:8c}
\\
& +\kappa \left( 1-h(r)^{2}\right) h(r)=0,            \nonumber
\end{flalign}
where the  dimensionless  parameter  $\kappa = \lambda e^{-2}$, and  the  prime
indicates  the derivative with respect to $\rho$.
Substituting ansatz (\ref{II:7}) into the expression for the $T_{00}$ component
of symmetric energy-momentum  tensor (\ref{II:6})  and  integrating it over the
space, we obtain the expression for the energy of the dyon
\begin{flalign}
E & = m_{M}\int\limits_{0}^{\infty }\biggl[ \frac{
u^{\prime }\left( \rho \right) ^{2}}{\rho ^{2}}+\frac{1}{2}\left( h^{\prime
}\left( \rho \right) ^{2}+j^{\prime }\left( \rho \right) ^{2}\right) \biggr.
\nonumber
\\
&+\left. \frac{\left( u\left( \rho \right) ^{2}-1\right) ^{2}}{2\rho ^{4}}+
\frac{\left( h\left( \rho \right) ^{2}+j\left( \rho \right) ^{2}\right)
u\left( \rho \right) ^{2}}{\rho ^{2}}\right.
\nonumber
\\
&+\left. \frac{\kappa}{4}\left(1 - h\left( \rho \right) ^{2}\right)^{2}
\right] \rho ^{2}d\rho,                                            \label{II:9}
\end{flalign}
where $m_{M} = 4 \pi v e^{-1}$ is the mass of the BPS monopole \cite{prs_1975}.

The  regularity  of  the   dyon   solution  at  $r = 0$  and  the finiteness of
dyon  energy  (\ref{II:9}) lead us to  the  boundary  conditions for the ansatz
functions:
\begin{subequations}                                              \label{II:10}
\begin{align}
&j(0) = 0, \qquad \underset{\rho\rightarrow \infty}{\lim}j(r)=c, \label{II:10a}
 \\
&u(0) = 1, \qquad \underset{\rho\rightarrow \infty}{\lim}u(r)=0, \label{II:10b}
 \\
&h(0) = 0, \qquad \underset{\rho\rightarrow \infty}{\lim}h(r)=1, \label{II:10c}
\end{align}
\end{subequations}
where $c$ is a finite value.
It follows  from  Eqs.~(\ref{II:8a})  and  (\ref{II:10})  that at large $\rho$,
the  ansatz  function $u(\rho) \infty \exp \bigl[ -\left(1 - c^{2}\right)^{1/2}
\rho \bigr]$.
Hence, the limiting value $c$ of the ansatz function $j(\rho)$ must satisfy the
condition $\left\vert c \right\vert < 1$.

The Higgs field  (\ref{II:7c})  is  not  invariant  under  the  initial $SU(2)$
gauge group of model (\ref{II:1}),  but it remains  invariant  under the $U(1)$
gauge  subgroup that  corresponds  to  local  rotations around  the unit vector
$n^{a} = x^{a}/r$ in the isospace.
This leads to the existence of the long-range gauge field that can be described
by the field strength  tensor $F_{\mu \nu } = v^{-1}\phi ^{a}F_{\mu \nu }^{a}$.
The corresponding intensities of the electric  and  magnetic fields of the dyon
are
\begin{equation}
E_{i}=v^{-1}F_{0i}^{a}\phi ^{a}=-ev^{2}j^{\prime }n_{i}           \label{II:11}
\end{equation}
and
\begin{equation}
B_{i} = \left(2 v\right)^{-1}\epsilon_{ijk}F_{jk}^{a}\phi^{a} = e v^{2}
\frac{1 - u^{2}}{\rho ^{2}}n_{i},                                 \label{II:12}
\end{equation}
respectively.
We define the electric (magnetic) charge  of  the dyon $Q_{E}$ ($Q_{M}$) as the
flux of the electric (magnetic) field  through  the  infinitely  distant sphere
$S^{2}_{\infty}$ and obtain the expressions
\begin{eqnarray}
Q_{E} &=&\oint_{S_{\infty }^{2}}d^{2}S_{n}E_{n}=-\frac{4\pi }{e}\underset{
\rho \rightarrow \infty}{\lim}\rho^{2}j^{\prime}\left(\rho\right) \nonumber
\\
&=&-\frac{8 \pi}{e} \int\nolimits_{0}^{\infty } j(\rho)
u(\rho)^{2} d\rho                                                 \label{II:13}
\end{eqnarray}
and
\begin{equation}
Q_{M}=\oint_{S_{\infty }^{2}}d^{2}S_{n}B_{n}=\frac{4\pi }{e}.     \label{II:14}
\end{equation}
To obtain the second line in Eq.~(\ref{II:13}), we use Gauss's law (\ref{II:8b})
written  in  the form $\left(\rho^{2}j^{\prime}\right)^{\prime} = 2 j u^{2}$.

The dyon's energy (\ref{II:9}) can be written as the sum of terms
\begin{equation}
E = E^{\left(E\right)} + E^{\left(B\right)} + E^{\left(G\right)}
+E^{\left(P\right)},                                              \label{II:15}
\end{equation}
where
\begin{flalign}
E^{\left( E\right) } &=m_{M}\int\limits_{0}^{\infty }\left[ \frac{1}{2}
j^{\prime }\left( \rho \right) ^{2}+\frac{u\left( \rho \right)^{2}j
\left(\rho \right) ^{2}}{\rho ^{2}}\right] \rho ^{2} d\rho         \nonumber
\\
&=-\frac{1}{2} v c Q_{E}                                          \label{II:16}
\end{flalign}
is the electric field's energy,
\begin{equation}
E^{\left(B\right) }=m_{M}\int\limits_{0}^{\infty }\left[ u^{\prime }\left(
\rho \right) ^{2}+\frac{(u\left( \rho \right) ^{2}-1)^{2}}{2\rho ^{2}}\right]
d\rho                                                             \label{II:17}
\end{equation}
is the magnetic field's energy,
\begin{equation}
E^{\left( G\right) }=m_{M}\int\limits_{0}^{\infty }\left[ \frac{1}{2}%
h^{\prime }\left( \rho \right) ^{2}+\frac{h\left( \rho \right) ^{2}u\left(
\rho \right) ^{2}}{\rho ^{2}}\right] \rho ^{2}d\rho               \label{II:18}
\end{equation}
is the gradient part of the soliton’s energy, and
\begin{equation}
E^{\left( P\right) }=m_{M}\int\limits_{0}^{\infty }\left[ \frac{1}{4}\kappa
\left( 1-h\left( \rho \right) ^{2}\right) ^{2}\right] \rho ^{2}d\rho
                                                                  \label{II:19}
\end{equation}
is the potential part of the soliton’s energy.
Any   solution   of   Eqs.~(\ref{II:8a}) -- (\ref{II:8c})  satisfying  boundary
conditions (\ref{II:10}) is a stationary point of the action $S= \int \mathcal{
L} d^{3}x dt$.
However, the  Lagrangian  density  (\ref{II:1})  does not depend on time if the
field configurations are those of ansatz (\ref{II:7}).
Hence, any solution of Eqs.~(\ref{II:8a}) -- (\ref{II:8c}) and (\ref{II:10}) is
a stationary point of the Lagrangian
\begin{equation}
L = \int \mathcal{L}d^{3}x = E^{\left(E\right)}-E^{\left(B\right)}-
E^{\left(G\right)}-E^{\left(P\right)}.                            \label{II:20}
\end{equation}

After the scale transformation $\rho \rightarrow \varkappa \rho$ of the argument
of the solution, the Lagrangian  $L$  becomes a function of the scale parameter
$\varkappa$.
Note  that  this  transformation  is  valid  because  Gauss's  law (\ref{II:8b})
remains  true  even  after  the  rescaling  $\rho \rightarrow \varkappa  \rho$.
Since the function $L (\varkappa)$ has a stationary point at $\varkappa=1$, its
derivative vanishes  at this point: $\left. dL/d\varkappa\right\vert_{\varkappa
= 1} = 0$.
It can easily be shown that $E^{\left( E\right) }\rightarrow \varkappa ^{-1}E^{
\left( E\right)}$, $E^{\left( B\right)}\rightarrow \varkappa E^{\left( B\right)
}$, $E^{\left( G\right) } \rightarrow \varkappa ^{-1}E^{\left( G\right) }$, and
$E^{\left( P\right)} \rightarrow \varkappa ^{-3}E^{\left( P\right)}$  under the
rescaling $\rho \rightarrow \varkappa \rho$.
Using this fact and  Eqs.~(\ref{II:16}) -- (\ref{II:20}),  we obtain the virial
relation for the dyon solution
\begin{equation}
E^{\left(E\right)} + E^{\left(B\right)} - E^{\left(G\right)} - 3 E^{\left(
P\right)} = 0.                                                    \label{II:21}
\end{equation}

In the important case of the BPS dyon \cite{prs_1975},  we  have the analytical
expressions for the energy components and the total energy:
\begin{subequations}                                              \label{II:22}
\begin{eqnarray}
E^{\left( E\right) } &=&\frac{v}{2}\frac{Q_{E}^{2}}{\sqrt{Q_{M}^{2}+Q_{E}^{2}
}},                                                              \label{II:22a}
\\
E^{\left( B\right) } &=&\frac{v}{2}\frac{Q_{M}^{2}}{\sqrt{Q_{M}^{2}+Q_{E}^{2}
}},                                                              \label{II:22b}
\\
E^{\left( G\right) } &=&\frac{v}{2}\sqrt{Q_{M}^{2}+Q_{E}^{2}},   \label{II:22c}
\\
E^{\left( P\right) } &=&0,                                       \label{II:22d}
\\
E &=&v\sqrt{Q_{M}^{2}+Q_{E}^{2}}.                                \label{II:22e}
\end{eqnarray}
\end{subequations}
It can easily be checked that Eqs.~(\ref{II:22}) satisfy  Eq.~(\ref{II:15}) and
virial relation (\ref{II:21}).

\section{A differential relation between the energy and electric charge and its
         consequences} \label{seq:III}

To obtain the differential relation between  the energy and electric charge, we
begin with the consideration of the BPS limit $\kappa = 0$.
In this  case,  there  is  the  analytical  solution  \cite{prs_1975} of system
(\ref{II:8a}) -- (\ref{II:8c})
\begin{flalign}
& u\left( \rho \right) =\frac{\tau\rho}{\sinh \left(\tau\rho \right)},
                                                                 \label{III:1a}
\\
& j\left( \rho \right)  =-\frac{Q_{E}}{Q_{M}}\tau\left[ \coth \left( \tau\rho
\right) -\left( \tau\rho \right) ^{-1}\right],                   \label{III:1b}
\\
& h\left( \rho \right)  =\coth \left( \tau\rho \right)-\left(\tau\rho \right)
^{-1},                                                           \label{III:1c}
\end{flalign}
where the ratio
\begin{equation}
\tau=\frac{Q_{M}}{\sqrt{Q_{M}^{2}+Q_{E}^{2}}}.                    \label{III:2}
\end{equation}
Eq.~(\ref{III:1b}) tells us that in the BPS case
\begin{equation}
c\equiv j\left( \infty \right) = -\frac{Q_{E}}
{\sqrt{Q_{M}^{2}+Q_{E}^{2}}}.                                     \label{III:3}
\end{equation}
Using Eqs.~(\ref{II:22e}) and (\ref{III:3}), we obtain successively
\begin{subequations}                                              \label{III:4}
\begin{eqnarray}
\frac{dE}{dc} &=&m_{M}c\left( 1-c^{2}\right) ^{-3/2},            \label{III:4a}
\\
\frac{dQ_{E}}{dc} &=&-\frac{m_{M}}{v}
\left(1-c^{2}\right)^{-3/2},                                     \label{III:4b}
\\
\frac{dE}{dQ_{E}} &=&\frac{dE/dc}{dQ_{E}/dc}=-vc,                \label{III:4c}
\end{eqnarray}
\end{subequations}
where $m_{M} = v Q_{M} = 4\pi v/e$ is the mass of the BPS monopole.
We see that in  the  BPS  case, the derivatives $dE/dc$ and $dQ_{E}/dc$ satisfy
the condition $dE/dc + v c dQ_{E}/dc = 0$.

Now we shall show that this condition is also valid in the general case $\kappa
\ne 0$.
To do this, we go from the ansatz function $j(\rho)$  to the new one $J(\rho) =
j(\rho) - c$ satisfying the  homogeneous  boundary condition $J(\infty) = 0$ at
infinity.
We consider ansatz functions (\ref{II:7}) as functions  of  the radial variable
$\rho$ and the parameter $c$.
Next, we calculate the value $dE/dc+ v c dQ_{E}/dc$ using Eqs.~(\ref{II:9}) and
(\ref{II:13}) for the energy and the electric charge, respectively.
The resulting expression contains, among  others, the two terms: $m_{M}\rho^{2}
\left(\partial h/\partial \rho\right) \left(\partial^{2}h/\partial\rho \partial
c\right)$ and $2m_{M}\left( \partial u/\partial \rho \right) \left(\partial^{2}
u/\partial\rho \partial c\right)$.
Using boundary  conditions (\ref{II:10}) and integration by parts, we transform
these two terms to $ -m_{M}\left[ 2\rho \left( \partial h/\partial \rho \right)
\left(\partial h/\partial c\right) +\rho^{2}\left( \partial ^{2}h/\partial \rho
^{2} \right) \left(\partial h/\partial c \right) \right]$  and  $ -2m_{M}\left(
\partial^{2} u/\partial \rho ^{2}\right) \left( \partial u/\partial c \right)$,
respectively.
After that, the  expression $dE/dc + v c dQ_{E}/dc$  can be written in the form
\begin{flalign}
&\frac{dE}{dc}+vc\frac{dQ_{E}}{dc}=-m_{M}\int \nolimits_{0}^{\infty }
\left( 2\frac{\partial u}{\partial c}e_{1}%
\right.                                                           \label{III:5}
\\
&\left. +\rho ^{2}J\frac{\partial e_{2}}{\partial c}
+\rho ^{2}\frac{\partial h}{\partial c}e_{3}-\frac{%
\partial }{\partial \rho }\left[ \rho ^{2}J\frac{\partial ^{2}J}{\partial
\rho \partial c}\right] \right) d\rho,                 \nonumber
\end{flalign}
where   $e_{1}$,  $e_{2}$,  and   $e_{3}$   are   the   left   hand   sides  of
Eqs.~(\ref{II:8a}), (\ref{II:8b}), and (\ref{II:8c}), respectively.
It is obvious that $e_{i}$ and  their derivatives with respect to the parameter
$c$ vanish  when  $u(\rho, c)$, $j(\rho, c)$, and $h(\rho, c)$ is a solution of
system (\ref{II:8a}) -- (\ref{II:8c}).
The last term $m_{M} \int\nolimits_{0}^{\infty} \partial \left[\rho^{2} J\left(
\partial^{2}J/\partial \rho \partial c\right)\right] /\partial \rho d\rho$ also
vanishes because $J\left(\partial^{2}J/\partial\rho\partial c\right) \sim\rho^{
-3}$ as $\rho \rightarrow \infty$.
We see that the  dyon's  energy  and  the electric charge satisfy  differential
relation (\ref{III:4c}) also in the general case $\kappa \ne 0$.
Eq.~(\ref{III:4c}) can be written in the form
\begin{equation}
\frac{dE}{dQ_{N}} \equiv e \frac{dE}{dQ_{E}} = \Omega,            \label{III:6}
\end{equation}
where $Q_{N} = e^{-1} Q_{E}$  is the Noether charge and the parameter $\Omega =
-evc=-m_{V}c$ is some function of $Q_{N}$.

The differential relation (\ref{III:6})  has  the same form as the differential
relations for the  nontopological  solitons in Refs.~\cite{gulamov_2014, fried,
fried1, fried2}. 
In the latter case, the differential  relation  results  from the fact that the
nontopological soliton is  a  stationary  point  of the total energy functional
under the condition  that  the Noether charge of field configurations is fixed.
A similar situation takes place for the dyon.
To  show  this,  we  give  an  interpretation  of  the  parameter  $\Omega$  in
Eq.~(\ref{III:6}).
Using  the  second  line  of  Eq.~(\ref{II:16}),  which  is  a  consequence  of
Gauss's law (\ref{II:8b}), we  can write the Lagrangian $L = E^{\left(E\right)}
- E^{\left(B\right)} - E^{\left(G\right)} - E^{\left(P\right)}$ in the form
\begin{equation}
L = \Omega Q_{N} - E.                                           \label{III:VII}
\end{equation}
If boundary conditions are fixed then variations of $L$ must vanish on the dyon
solution of field equations.
The fixation of the boundary conditions means that the parameter $\Omega = -evc
= -evj\left( \infty \right)$ remains fixed when varying the Lagrangian $L$, and
hence
\begin{equation}
-\delta L = \delta E - \Omega \delta Q_{N} = 0.                \label{III:VIII}
\end{equation}
It follows from  Eq.~(\ref{III:VIII})  that  the  dyon solution is a stationary
point of the total energy functional $E$  provided that  the Noether (electric)
charge $Q_{N}$ ($Q_{E}$) of  field  configurations  is fixed, and the parameter
$\Omega$ plays the  role  of  the  Lagrange multiplier in Eq.~(\ref{III:VIII}).

We now turn to the unitary gauge $\boldsymbol{\phi} = (0,0,\chi)$ and shall use
the Hamiltonian formalism.
In the unitary gauge, the canonical fields are
\begin{equation}
A_{i}^{a},\;P_{i}^{a} = F_{0i}^{a} = E_{i}^{a},\;\chi,\;p = \partial_{t}\chi,
                                                                 \label{III:IX}
\end{equation}
and the Hamiltonian density is
\begin{flalign}
\mathcal{H} &=\frac{1}{2}P_{i}^{a}P_{i}^{a}+\frac{1}{2}p^{2}+\frac{1}{4}
F_{ij}^{a}F_{ij}^{a}                                          \nonumber
\\
&+\frac{1}{2}e^{2}\chi ^{2}\left(
A_{0}^{1}A_{0}^{1}+A_{i}^{1}A_{i}^{1}+A_{0}^{2}A_{0}^{2}+A_{i}^{2}A_{i}^{2}
\right)                                                       \nonumber
\\
&+\frac{1}{2}\left( \nabla \chi \right) ^{2}+\frac{\lambda }{4}\left( \chi
^{2} - v^{2}\right)^{2}.                                          \label{III:X}
\end{flalign}
Gauss's law (the zeroth  component  of  Eq.~(\ref{II:5a})  written  in terms of
canonical fields (\ref{III:IX}))
\begin{flalign}
A_{0}^{a}-\delta _{3}^{a}A_{0}^{3} &=-\left( e^{2}\chi ^{2}\right)
^{-1}\left( \partial _{i}P_{i}^{a}
-e\epsilon^{abc}A_{i}^{b}P_{i}^{c}\right)                     \nonumber
\\
&=-\left( e^{2}\chi ^{2}\right)^{-1}D_{i}P_{i}^{a}               \label{III:XI}
\end{flalign}
is used to express  $A_{0}^{1,2}$  in  Eq.~(\ref{III:X}).
Note that the boundary condition  $A^{3}_{0}(\infty)=0$, which completely fixes
the unitary gauge, is used to obtain Eq.~(\ref{III:XI}) within the framework of
the Hamiltonian formalism.

In the unitary gauge, the unbroken electromagnetic  $U(1)$ subgroup corresponds
to rotations about the third axis in the isospace.
The corresponding Noether charge
\begin{equation}
Q_{N} = e^{-1}Q_{E} = \int\epsilon^{3bc}A_{i}^{b}P_{i}^{c}d^{3}x \label{III:XII}
\end{equation}
is consistent with the definition  of  Eq.~(\ref{II:13}) because  of  the third
component of Gauss's law (\ref{III:XI}) and the definition $E_{i}\equiv E_{i}^{
3} = P_{i}^{3}$ following from Eq.~(\ref{III:IX}).
Further, it can be shown that
\begin{equation}
H = \int \mathcal{H}d^{3}x = \int T_{00}d^{3}x = E             \label{III:XIII}
\end{equation}
for field configurations satisfying Gauss's law (\ref{III:XI}).
Now we use the trick  used in Refs. \cite{fried1, fried2, saf1}.
Taking   into     account     Eqs.~(\ref{III:VIII}),     (\ref{III:XII}),   and
(\ref{III:XIII}), the Hamilton equations can be written in the form
\begin{flalign}
\partial _{t}A_{i}^{a} &= \frac{\delta H}{\delta P_{i}^{a}}=\frac{\delta E}{
\delta P_{i}^{a}}=\Omega \frac{\delta Q_{N}}{\delta P_{i}^{a}} = \Omega
\epsilon ^{a3b}A_{i}^{b},                                      \label{III:XIVa}
\\
\partial _{t}P_{i}^{a} &= -\frac{\delta H}{\delta A_{i}^{a}}=-\frac{\delta E
}{\delta A_{i}^{a}}=-\Omega \frac{\delta Q_{N}}{\delta A_{i}^{a}} \nonumber
\\
& =\Omega \epsilon ^{a3b}P_{i}^{b}.                            \label{III:XIVb}
\end{flalign}
From Eq.~(\ref{III:XIVa}) it is  easy  to  obtain  the  time  dependence of the
fields $W_{\mu }^{\pm } = 2^{-1/2}\left( A_{\mu}^{1} \mp iA_{ \mu }^{2}\right)$
corresponding to electrically charged vector bosons:
\begin{equation}
W_{\mu }^{\pm }\left( t,\mathbf{x}\right) = \exp \left( \mp i\Omega t\right)
W_{\mu }^{\pm }\left( \mathbf{x}\right),                         \label{III:XV}
\end{equation}
whereas the remaining canonical fields  do  not  depend  on time in the unitary
gauge.
Thus, we conclude that the  parameter $\Omega$ entering in Eq.~(\ref{III:6}) is
the phase frequency of the charged vector boson fields in the unitary gauge.

Eqs.~(\ref{II:22}) -- (\ref{III:3})  tell  us  that  the energy and the Noether
(electric) charge of the BPS dyon increase  indefinitely as $\Omega \rightarrow
m_{V}$.
At the same time, it was  shown numerically in Refs.~\cite{bkt_1999, bhkt_1999}
that in the non-BPS  case, the energy  and  electric charge of the dyon  cannot
exceed the maximum allowable values,  which  depend  on the model's parameters.
Let us show the impossibility  of  arbitrary large values for the dyon's energy
and electric charge in  the  non-BPS  case $\kappa \neq 0$  using  differential
relation (\ref{III:6}).
For this, we differentiate Eq.~(\ref{II:15}) with respect to $\Omega$.
Taking  into  account  Eqs.~(\ref{II:16})  and  (\ref{III:6}),  we  obtain  the
relation
\begin{flalign}
\frac{\Omega }{2}& =\frac{Q_{N}}{2}\frac{d\Omega }{dQ_{N}}+\frac{
dE^{\left( B\right) }}{dQ_{N}}+\frac{dE^{\left( G\right) }}{dQ_{N}}+\frac{
dE^{\left( P\right) }}{dQ_{N}}.                                   \label{III:7}
\end{flalign}

Let us suppose that in the non-BPS  case,  the energy and the Noether charge of
the dyon tend to infinity as $\Omega \rightarrow m_{V}$.
It can be shown that in this case, the term $Q_{N}\left(d\Omega /dQ_{N}\right)$
must tend to zero.
To do this,  we  write  the  differential  equation $Q_{N}\left(d\Omega /dQ_{N}
\right)=F\left(Q_{N}\right)$ and integrate it:
\begin{equation}
\Omega \left( Q_{N}\right) =\Omega \left( \bar{Q}_{N}\right)
+\int\nolimits_{\bar{Q}_{N}}^{Q_{N}}\bar{\bar{Q}}_{N}^{-1}
F\bigl(\bar{\bar{Q}}_{N}\bigr) d\bar{\bar{Q}}_{N}.                \label{III:8}
\end{equation}
Because $\underset{Q_{N}\rightarrow \infty }{\lim } \Omega \left(Q_{N}\right) =
m_{V}$, the integral on the right side of Eq.~(\ref{III:8}) must  remain finite
as $Q_{N}\rightarrow \infty$.
This is  only  possible  if $F\left( Q_{N}\right) = Q_{N}\left( d\Omega /dQ_{N}
\right)$  tends  to zero as $Q_{N} \rightarrow \infty$.

Next, we  estimate  the  derivative  $d E^{(B)}/d Q_{N}$  as $Q_{N} \rightarrow
\infty$.
By analogy with electrostatics,  the energy of the dyon's magnetic field can be
written in the form $E^{\left(B \right)} = Q_{M}^{2}/\left( 8\pi R_{M}\right)$,
where $Q_{M} = 4\pi/e$ is the dyon's magnetic charge  and $R_{M}$ is the dyon's
effective magnetic radius.
With the  increase  in  the  Noether  (electric)  charge $Q_{N}$ ($Q_{E}$), the
effective size of the dyon also increases  as  in  the  BPS case (\ref{III:1a})
 -- (\ref{III:2}).
Hence, the  effective  magnetic  radius  $R_{M}$  also increases (or, at least,
remains bounded from below) when $Q_{N} \rightarrow \infty$.
It follows that the magnetic field's energy $E^{\left( B\right)}(Q_{N})=Q_{M}^{
2}/\left(8\pi R_{M}(Q_{N})\right)$ is a bounded function on the interval $Q_{N}
\in \left[ 0,\infty \right)$.
But the derivative of a function bounded  on  the semi-infinite interval $Q_{N}
\in \left[ 0,\infty \right)$ must tend  to  zero as $Q_{N} \rightarrow \infty$.
The only exception is an oscillating function of $Q_{N}$, but  it is clear that
this variant cannot be realised.
Thus, we  conclude  that  $dE^{\left( B\right) }/dQ_{N}\rightarrow 0$ as $Q_{N}
\rightarrow \infty$.

Next, we consider the behaviour of the derivatives $dE^{\left(G\right)}/dQ_{N}$
and $dE^{\left(P\right)}/dQ_{N}$ as $Q_{N} \rightarrow \infty$.
For  this, we  rewrite  Eq.~(\ref{II:18})  integrating  by  parts the term $h^{
\prime 2}/2$ and use Eq.~(\ref{II:8c}) to obtain
\begin{equation}
E^{\left( G\right) }=m_{M} \frac{\kappa}{2} \int\limits_{0}^{\infty }\left[
\left(1 - h\left( \rho \right) ^{2}\right) h\left( \rho \right) ^{2}\right]
\rho^{2}d\rho.                                                    \label{III:9}
\end{equation}
Note that Eq.~(\ref{III:9}) is  valid  only  for  the  non-BPS case because the
integral diverges in the BPS case.
Using Gauss's law (\ref{II:8b}),  it  can easily be shown that $j(\rho)$ is
a monotonic function on the interval $\rho \in \left[ 0, \infty  \right)$.
Furthermore, boundary  conditions  (\ref{II:10a}) tell  us  that $j(\rho)$ is a
bounded function.
Then, it follows from Eq.~(\ref{II:13}) that  the integral $\int\nolimits_{0}^{
\infty} u\left( \rho \right) ^{2}d\rho$ diverges as $Q_{N} \rightarrow \infty$.
Eq.~(\ref{II:8a}) tells us that this is only possible if, on the arbitrary large
interval $\rho \in \left[0, \varrho \right]$,  the function $u(\rho)$ is in the
infinitesimal neighbourhood of $1$ and $h(\rho) \approx j(\rho)$.
It follows from Eq.~(\ref{II:8b}) that, in this case, the function $j\left(\rho
\right)\approx \bar{c}\rho$, where $\bar{c}$ is a constant.
For $\rho\gtrsim\varrho$, the function $u(\rho)\approx 0$ and Eq.~(\ref{II:8b})
together with boundary condition (\ref{II:10a}) tell us that $j\left(\rho\right)
\approx c + e^{2}Q_{N}/(4\pi \rho)$ in this case.
We estimate the constant $\bar{c}$ by equating the two expressions for $j\left(
\rho\right)$ at $\rho=\varrho$ and obtain that $\bar{c}=c/\left(2\varrho\right)
= -\Omega/(2 m_{V} \varrho)$.
Using this expression and Eq.~(\ref{II:13}),  we  obtain the leading asymptotic
behaviour of the Noether charge $Q_{N}=e^{-1}Q_{E}$
\begin{equation}
Q_{N}\approx \frac{2\pi \varrho \Omega }{e^{2}m_{V}}.            \label{III:10}
\end{equation}

We now have everything we need  to  find  the  leading  asymptotic behaviour of
$E^{\left( G\right) }$  and $E^{\left(P\right)}$ as $Q_{N} \rightarrow \infty$.
Using Eqs.~(\ref{II:19}),  (\ref{III:9}),  and  (\ref{III:10})  and considering
that $h(\rho)\approx j(\rho)$ when $\rho \in \left[0,\varrho\right]$, we obtain
the expressions: 
\begin{equation}
E^{\left(P\right)} \approx 2.95\,E^{\left(G\right)} \approx 0.000244\, m_{M}
\kappa e^{6}Q_{N}^{3}.                                           \label{III:11}
\end{equation}
We see that both $E^{\left(P\right)}$ and $E^{\left(G\right)}$ are $\propto Q_{
N}^{3}$ in the limit of large $Q_{N}$.
It follows that the  derivatives $dE^{\left( P\right) }/dQ_{N}$ and $dE^{\left(
G\right) }/dQ_{N}$   are   $\propto Q_{N}^{2}$,  and  hence  diverge  as $Q_{N}
\rightarrow \infty$.
But this is incompatible with Eq.~(\ref{III:7}) (and therefore with differential
relation (\ref{III:6})), which implies that the derivatives $dE^{\left(P\right)
}/dQ_{N}$ and $dE^{\left(G\right)}/dQ_{N}$ must be  finite  because $\Omega \in
\left( -m_{V}, m_{V} \right)$.
It is obvious that virial  relation  (\ref{II:21})  also cannot be satisfied in
this case.
It follows that the energy and  Noether (electric) charge of the dyon cannot be
arbitrarily large in the non-BPS case.

In the BPS case, the  potential  part $E^{(P)}$  of the dyon's energy is absent
and Eq.~(\ref{III:9})  becomes  inapplicable  along  with  our  conclusion that
$E^{(G)} \propto Q_{N}^{3}$ in the limit of large $Q_{N}$.
Here we need to use Eq.~(\ref{II:18}) in order to find that $E^{\left( G\right)
} \rightarrow m_{V}Q_{N}/2$  as  $Q_{N} \rightarrow \infty$  in accordance with
Eq.~(\ref{II:22c}).
Furthermore, using  Eqs.~(\ref{II:22}) it  can  be  shown  that the combination
$\left( Q_{N}/2\right)\left(d\Omega /dQ_{N}\right) +dE^{\left(B\right)}/dQ_{N}$
vanishes in the BPS case and Eq.~(\ref{III:7}) takes the form
\begin{equation}
\frac{dE^{\left( G\right) }}{dQ_{N}}=\frac{\Omega }{2}.          \label{III:12}
\end{equation}
The  correctness   of   Eq.~(\ref{III:12})  can   be   easily   verified  using
Eqs.~(\ref{II:22}).
It  follows  that  Eqs.~(\ref{III:6})  and  (\ref{III:7})  do  not  impose  any
restrictions in the BPS case, and  therefore  the energy and Noether (electric)
charge of the BPS dyon can be arbitrarily large as $\Omega \rightarrow m_{V}$.

\section{Numerical results}
\label{seq:IV}

Now we  present  some  numerical  results  concerning the dyon.
For numerical calculations,  we  use the natural units $c = 1$ and $\hbar = 1$.
It follows from  Eqs.~(\ref{II:8a}) -- (\ref{II:8c}) and (\ref{II:10}) that the
ansatz functions $u(\rho)$, $j(\rho)$, and $h(\rho)$ depend  only  on  the  two
dimensionless  parameters  $\kappa = e^{-2}\lambda$  and $c = -m_{V}^{-1}\Omega
\equiv -\tilde{\Omega}$.
Further, Eqs.~(\ref{II:9}) and (\ref{II:13}) tell  us  that  the  dimensionless
combinations $\tilde{Q}_{E} = eQ_{E}$ and $\tilde{E} = e^{2} E m_{V}^{-1}$ also
depend only on $\kappa = e^{-2}\lambda$ and $\tilde{\Omega}= m_{V}^{-1}\Omega$.

\begin{figure}[t]
\includegraphics[width=7.8cm]{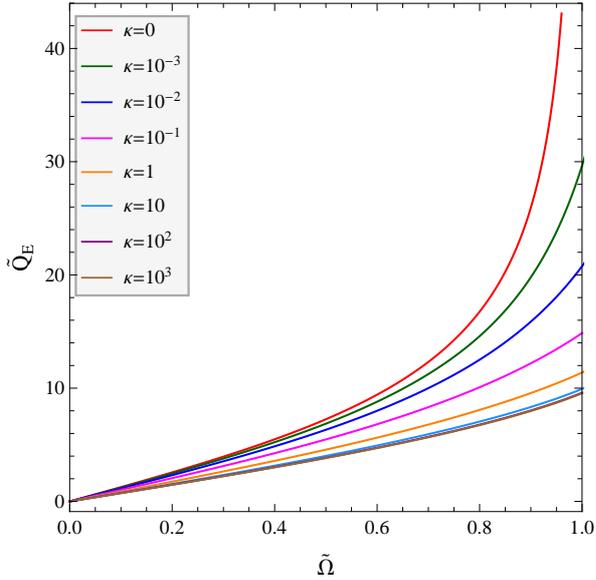}
\caption{\label{fig1}  Dependence of the dimensionless combination $\tilde{Q}_{
E} = e Q_{E}$ on $\tilde{\Omega}=m_{V}^{-1}\Omega$  for different values of the
parameter $\kappa = e^{-2} \lambda$.}
\end{figure}

\begin{figure}[t]
\includegraphics[width=7.8cm]{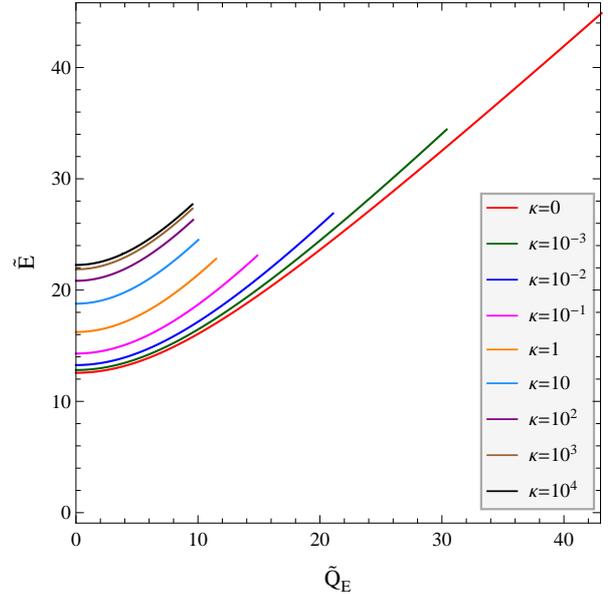}
\caption{\label{fig2}  Dependence of the dimensionless combination $\tilde{E} =
e^{2} E m_{V}^{-1} $ on $\tilde{Q}_{E} = e Q_{E}$  for different values  of the
parameter $\kappa = e^{-2} \lambda$.}
\end{figure}

Figure~\ref{fig1}  presents   the   dependence   of   $\tilde{Q}_{E}$   on  the
dimensionless  phase  frequency  $\tilde{\Omega}$ for  different  values of the
parameter $\kappa = e^{-2} \lambda$.
We see that for any $\kappa$, the dyon's electric charge increases monotonically
with an increase in $\Omega$.
The electric charge of  the BPS dyon ($\kappa = 0$) tends to infinity as $Q_{E}
\sim 2\sqrt{2} \pi e^{-1}m_{V}^{1/2}\left( m_{V} - \Omega \right)^{-1/2}$  when
$\Omega \rightarrow m_{V}$, in accordance with Eq.~(\ref{III:3}).
At the same time, the electric charge  of  the  non-BPS  dyon remains finite as
$\Omega \rightarrow m_{V}$.
For any fixed $\Omega$, the dyon's electric charge decreases monotonically with
an increase in $\kappa$ and tends to some limiting value as $\kappa \rightarrow
\infty$.

Figure~\ref{fig2} presents the dependence  of  the  dimensionless scaled energy
$\tilde{E}$ on the scaled electric charge $\tilde{Q}_{E}$ for  different values
of the parameter $\kappa = e^{-2} \lambda$.
It follows from Fig.~\ref{fig2} that for any  fixed $\kappa$, the dyon's energy
increases monotonically with an increase in the electric charge $Q_{E}$.
Further, for any nonzero $\kappa$,  there  exist the maximum permissible values
for the dyon's energy $E$ and electric charge $Q_{E}$.
They correspond to the rightmost points on the curves in Fig.~\ref{fig2}, where
the derivative $d\tilde{E}/d\tilde{Q}_{E} = 1$  according to Eq.~(\ref{III:6}).
For any  fixed  $Q_{E}$,  the  dyon's  energy increases monotonically   with an
increase in  $\kappa$,  but  the  maximum  allowable energy depends on $Q_{E}$.
As $\kappa \rightarrow \infty$,  the  curves $\tilde{E}(\tilde{Q}_{E})$ tend to
the limiting curve.
Note that the  dyon's  energy  $E$  is  an even function of the electric charge
$Q_{E}$ due to the $C$-invariance  of  model (\ref{II:1}).

It follows from Fig.~\ref{fig2}  that  for  any  fixed  $\kappa$,  the function
$E(Q_{E})$  is  a  convex  downward,  and   therefore   the  second  derivative
$d^{2}E/d Q_{N}^{2}$ is positive.
Hence, the function $E(Q_{E})$ satisfies the inequality
\begin{equation}
E\left( Q_{E}\right) \leq E\left( Q_{E}^{\prime }\right) +m_{V}e^{-1}\left(
Q_{E}-Q_{E}^{\prime }\right).                                      \label{IV:1}
\end{equation}
It follows that the dyon having  the  electric charge $Q_{E}$ is stable against
decay into the  the  dyon  with  the  smaller  electric charge $Q_{E}^{\prime}$
and the massive  gauge  bosons  with the total electric charge $Q_{E} - Q_{E}^{
\prime}$  and mass $m_{V}e^{-1}(Q_{E} - Q_{E}^{\prime}) = m_{V}(Q_{N} - Q_{N}^{
\prime})$.

\begin{figure}[t]
\includegraphics[width=7.8cm]{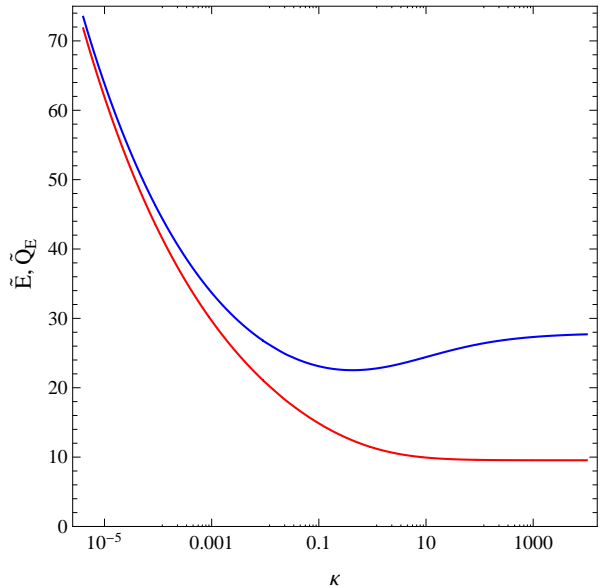}
\caption{\label{fig3}  Dependence of the maximum possible values of $\tilde{E}$
(upper line) and $\tilde{Q}_{E}$ (lower line) on the parameter $\kappa = e^{-2}
\lambda$.}
\end{figure}

In Fig.~\ref{fig3}, we can see  the  dependence  of the maximum possible values
of the dimensionless scaled energy $\tilde{E}$  and  the scaled electric charge
$\tilde{Q}_{E}$  on  the parameter $\kappa = e^{-2}\lambda$.
For better visualisation  of  the  $\kappa$-dependences,  we  show  them on the
log-linear plot.
It follows from  Fig.~\ref{fig3} that  the  maximum allowable value of $\tilde{
Q}_{E}$ decreases monotonically with an increase in $\kappa$ in accordance with
Ref.~\cite{bkt_1999}.
At the same time, the  maximum  allowable  value  of $\tilde{E}$ also decreases
with an increase in $\kappa$ until it reaches the minimum  point with $(\kappa,
\tilde{E}) = (0.418, 22.532)$; after this, it  increases  with  an  increase in
$\kappa$.
As $\kappa \rightarrow 0$, the  maximum  allowable  values  of  $\tilde{E}$ and
$\tilde{Q}_{E}$ increase indefinitely according to the power law:
\begin{equation}
\tilde{E} \sim \tilde{Q}_{E} \sim \alpha \kappa^{-\beta},          \label{IV:2}
\end{equation}
where $\alpha$ is a positive constant and $\beta \approx 0.165$.
When  $\kappa  \rightarrow  \infty$,  these   maximum   allowable  values  tend
asymptotically to the finite limits:
\begin{flalign}
\tilde{E}&\underset{\kappa \rightarrow \infty }{\longrightarrow}
\tilde{E}(\infty)  \approx  27.704,                               \label{IV:3a}
\\
\tilde{Q}_{E}&\underset{\kappa \rightarrow \infty}
{\longrightarrow} \tilde{Q}_{E}(\infty)  \approx  9.546.          \label{IV:3b}
\end{flalign}

Let us consider the limit in which  the  self-interaction constant $\lambda$ is
fixed and the gauge coupling constant $e$ tends to zero.
It follows that the combination $\kappa = e^{-2}\lambda$ increases indefinitely
in this limit.
Recalling the definition  of  $\tilde{E}$  and  $\tilde{Q}_{E}$ and taking into
account limiting values (\ref{IV:3a}) and (\ref{IV:3b}), we  obtain the leading
asymptotic behaviour of  the  dyon's  energy $E$ and electric charge $Q_{E}$ in
the limit $e \rightarrow 0$:
\begin{flalign}
E &\sim \tilde{E}(\infty) v e^{-1},                               \label{IV:4a}
\\
Q_{E} &\sim \tilde{Q}_{E}(\infty)e^{-1}.                          \label{IV:4b}
\end{flalign}
It follows that the dyon's  energy  and  electric  charge increase indefinitely
when $\lambda$ is fixed and $e \rightarrow 0$.

\section{Conclusions}
\label{seq:V}

In the present paper,  we have obtained the differential relation (\ref{III:6})
between the energy $E$ and electric charge $Q_{E}$ of the dyon solution.
The differential relation expresses the derivative $dE/dQ_{E}$  in terms of the
boundary value  for  the  temporal  component  of  the  dyon's  electromagnetic
potential.
Using the Hamiltonian formalism, it  is  shown  that, in the unitary gauge, the
derivative $dE/dQ_{E}$ is proportional to the phase rotation  frequency  of the
electrically charged  boson fields of the dyon.
It follows from the differential relation  that  the dyon is a stationary point
of the total energy functional  provided  that the Noether (electric) charge of
field configurations is fixed.
The latter property is a characteristic feature of the nontopological solitons,
and we therefore  conclude  that  the  dyon  possesses  the  properties of both
topological and nontopological solitons.

The differential  relation  (\ref{III:6}) results  in  the  boundedness  of the
derivative $dE/dQ_{E}$.
It follows that the energy and  the  electric charge of the non-BPS dyon cannot
be arbitrarily large.
The numerical study reveals that the dyon is stable against decay into the dyon
of smaller electric charge and massive electrically charged gauge bosons.
It also shows that the dyon's energy  and electric charge increase indefinitely
when the gauge coupling constant tends to zero.

\section*{Acknowledgements}

This work was supported by the Russian Science Foundation, grant No 19-11-00005.





\bibliographystyle{elsarticle-num}

\bibliography{article}






\end{document}